\documentclass[acmsmall, nonacm]{acmart}
\usepackage{nicefrac}
\usepackage[normalem]{ulem}

\AtBeginDocument{%
  }

\setcopyright{acmcopyright}
\copyrightyear{2023}
\acmYear{2023}
\acmDOI{XXXXXXX.XXXXXXX}

\begin{document}

\title{Towards a Deep(er) Understanding of Interaction through Modeling, Simulation, and Optimization}

\author{Florian Fischer}
\authornote{All authors contributed equally to this research.}
\email{florian.j.fischer@uni-bayreuth.de}
\orcid{0000-0001-7530-6838}

\author{Arthur Fleig}
\authornotemark[1]
\email{arthur.fleig@uni-bayreuth.de}
\orcid{0000-0003-4987-7308}

\author{Markus Klar}
\authornotemark[1]
\email{markus.klar@uni-bayreuth.de}
\orcid{0000-0003-2445-152X}

\author{Viktorija Paneva}
\authornotemark[1]
\email{viktorija.paneva@uni-bayreuth.de}
\orcid{0000-0002-5152-3077}

\author{Jörg Müller}
\email{joerg.mueller@uni-bayreuth.de}
\orcid{0000-0002-4971-9126}

\affiliation{%
  \institution{University of Bayreuth}
  \streetaddress{Universitätsstr.~30}
  \city{Bayreuth}
  \country{Germany}
  \postcode{D-95447}
}

\begin{abstract}
	The traditional user-centered design process can hardly keep up with the ever faster technical development and increasingly diverse user preferences.
	As a solution, we propose to augment the tried-and-tested approach of conducting user studies with simulation and optimization of the entire human-computer interaction loop. This approach allows to better understand phenomena through explicit modeling, build virtual prototypes through simulation, and improve interaction techniques through optimization.
	Building predictive user models also supports the creation and validation of HCI theories, and constitutes a decisive step towards new, intelligent, and adaptive user interfaces.
	We report our experience in virtually developing new interaction techniques on the example of acoustic levitation, and present our optimization-based framework for HCI. 
	With this, we strive to gain a better understanding of interaction and at the same time feed the discussion on questions such as which tools and tutorials are necessary to make virtual prototyping more accessible to different audiences. 
\end{abstract}



\maketitle

\section{Introduction}

As society becomes increasingly technologized, new tracking and display technologies, such as virtual, augmented, and mixed reality, have enabled massive growth in the design space of interaction techniques. 
However, interaction with technology still often feels lackluster and unnatural compared to how we interact with objects in nature.
To design more natural interaction techniques, it is crucial to better understand interaction techniques and user intentions. 
This becomes increasingly more challenging, as not only the devices but also user preferences become more diverse. 

The traditional user-centered design process in Human-Computer Interaction (HCI) is focused on creating interfaces that work well for a specific user group.
It tends to rely heavily on user feedback, which can be time-consuming, costly, and may not always provide actionable insights.
As such, it is struggling to keep up. 
We propose to add into the mix a principled model-based design process, where simulation and optimization of the whole human-computer interaction loop is key.
This does not replace the tried-and-tested approach of conducting user studies, but rather combines model-based simulation and optimization with user studies that need to be run much less frequently and at a much later stage.

Simulation provides a virtual environment for testing the behavior and performance of HCI systems and interfaces. 
It allows designers to keep the prototype entirely virtual, and evaluate user experience and assess user engagement, for example, in Virtual Reality (VR).
This helps to identify potential problems and to make improvements early on in the design process, before real prototypes are even built. 
Modeling the underlying interface dynamics is not only useful for the interaction design and testing, but also for optimizing the performance of the interface itself.

In addition to modeling an interaction technique, we can build a generative user model. 
This helps in identifying potential barriers to usability and ``makes design and engineering more predictable and robust processes''~\cite{interact}.
By simulating the user, we avoid physical, emotional, or ethical risks, and prevent causing stress to real users in exhausting user studies.
Finally, the ability of a simulation to match real user behavior is a strong indicator of whether we understand an interactive system, thus, building user models also supports the creation and validation of new theories in HCI.

Below we report on our most recent advances in this domain.
In Section 2, we investigate the development of future interactive systems for acoustic levitation interfaces, and provide examples how modeling and simulation can help assess user engagement and
test interaction parameters with virtual prototypes,
and improve interface performance.
In Section~\ref{sec:biomechanical-modeling}, we provide a comprehensive framework that encompasses the (biomechanical) simulation of users, where we focus on generating user \emph{movement}.
In Section~\ref{sec:discussion}, we reflect on limitations and current open questions in the field. 

\section{Acoustic Levitation}\label{sec:acoustic-levitation}

\begin{figure}[b]
    \centering
    \includegraphics[width=1\textwidth]{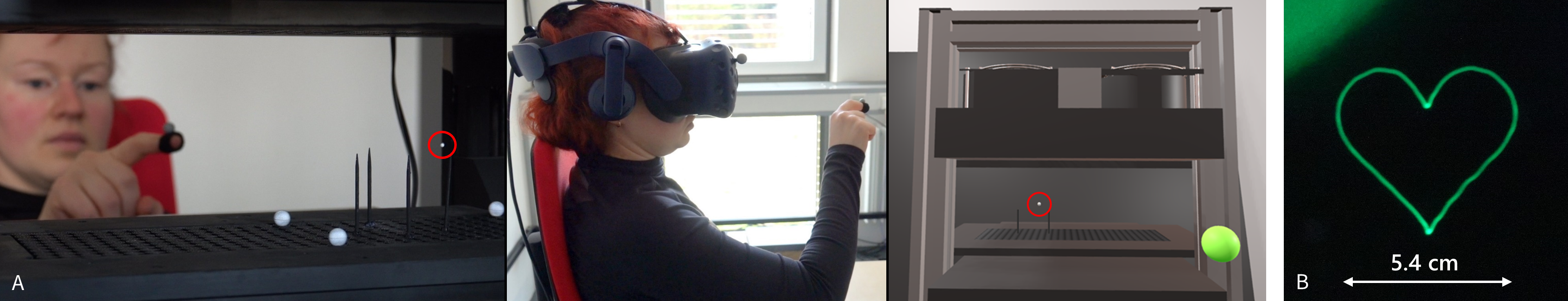}
    \caption{(A) A user performing a pointing task on the real prototype (left), and in the \textit{Levitation Simulator}~\cite{Paneva20} (middle, right), using  \textit{LeviCursor}~\cite{Bachynskyi18} for selecting targets in 3D space (marked in red). (B) A levitated particle traverses a periodic path at a frequency of 10Hz to reveal volumetric images in mid-air. Using the \textit{OptiTrap}~\cite{Paneva22} algorithm we can specify physically feasible trap trajectories that render generic shapes in optimal time.
 	Figure adapted from~\cite{Paneva20} and~\cite{Paneva22}.
    }
    \label{fig:Acoustic_Lev}
\end{figure}

Some of the greatest visionaries in HCI imagined the interface of the future as ``a room where the computer can control the existence of matter''~\cite{Sutherland65}, and as ``a dynamic physical material that reflects the changes in digital states in real time''~\cite{Ishii12}.
With acoustic levitation technology, we see vast potential to get a step closer to these great visions of the ultimate mixed reality, where the digital and the physical world are fully merged.
Acoustic levitation displays offer a novel and innovative way of displaying and interacting with digital content in real physical space, by using sound waves to manipulate physical matter.
The interface typically consists of two opposing phased arrays of transducers.
Each transducer emits ultrasonic waves at a frequency inaudible for humans, of 40kHz.
By appropriately setting the phase and amplitude of each transducer, we can generate \textit{acoustic traps} at the points where the acoustic forces converge.
In these nodes of low acoustic pressure, it is possible to suspend millimeter-sized particles in mid-air.
By moving the acoustic trap in 3D space, we can digitally control the position of the levitated physical matter, and in this manner, generate dynamic visualizations in physical space without the need for wearables or any other gadgets.

When developing and designing for radically novel interfaces such as acoustic levitation, in addition to the financial and time costs associated with user testing mentioned earlier, other challenges, such as availability and operability occur.
Building and maintaining interactive levitation interfaces requires specific technical expertise, knowledge of acoustics, microsecond synchronization, and submillimeter calibration of the system components. 
In addition, the interface can be difficult to debug. 
In case of problems, the only observable effect is the levitating particle shooting out in an uncontrollable manner. 
This can pose a barrier for designers, artists, game developers, researchers, etc.~to start experimenting with this novel technology, and test prototypes of their applications with users.
To solve this problem we propose virtual prototyping, an approach that has proven successful in other disciplines, e.g., automotive, product design, and manufacturing~\cite{GomesDeSa99, Berg17}. 
We developed the \textit{Levitation Simulator}~\cite{Paneva20} -- an interactive simulation tool in VR that can be used to iteratively develop and prototype ideas for acoustic levitation interfaces and even conduct user tests and formal experiments. 
Only after the development has converged, the resulting system can be validated using a real apparatus.
The simulator consists of two modules -- an interaction and a simulation module.
The interaction module is implemented within the Unity Game engine, and it can receive user input via a motion capture system or VR controllers. 
In the simulation module, we incorporated a model of a levitated particle moving in an acoustic field, which allowed for the simulation of physically accurate dynamics of the virtual particle. 
We validated the tool by performing a pointing study in the \textit{Levitation Simulator} and on the real prototype, using \textit{LeviCursor}~\cite{Bachynskyi18}, a levitated 3D physical cursor.
Figure~\ref{fig:Acoustic_Lev}(A) shows a user performing repetitive aimed movements in mid-air between two three-dimensional spherical targets on the real prototype, and in the \textit{Levitation Simulator}.
The results showed comparable performance. 
Further testing with gaming applications showed that the \textit{Levitation Simulator} can provide good predictions, regarding user interaction and engagement with the real prototype.
In future studies, the \textit{Levitation Simulator} can also be useful in exploring the multimodality of acoustic levitation displays, e.g., by augmenting the levitated particles with mid-air haptic feedback to potentially improve display accessibility~\cite{Paneva20haptiread, Carter13}.

Modeling of the underlying interface dynamics is not only useful for the interaction design and testing, but also for optimizing the performance of the  interface itself. 
We demonstrated this on an acoustic levitation interface that uses the persistence of vision effect to render smooth levitated graphics in real time, by rapidly moving a levitated particle along a periodic path. 
\textit{OptiTrap}~\cite{Paneva22} is an automated numerical approach that computes trap trajectories, i.e., position and timings of the acoustic traps, to generate physically feasible, nearly time-optimal paths that reveal generic mid-air shapes on the levitator. 
To achieve this, we derived a multi-dimensional model of the acoustic forces around a trap, and formulated and solved a non-linear path following problem.
As a result of this trajectory optimization, we were able to render bigger and more complex shapes (e.g., involving sharp edges and sudden changes in curvature), than previously possible (see Figure~\ref{fig:Acoustic_Lev}(B)).

On the example of acoustic levitation interfaces, we have demonstrated in practice, that modeling, simulation, and optimization methods can be of great benefit for the efficient and agile development of innovative user interfaces. 

\section{Modeling Human Movement During Interaction }\label{sec:biomechanical-modeling}

\begin{figure}[h!]
	\centering
	\includegraphics[width=\textwidth]{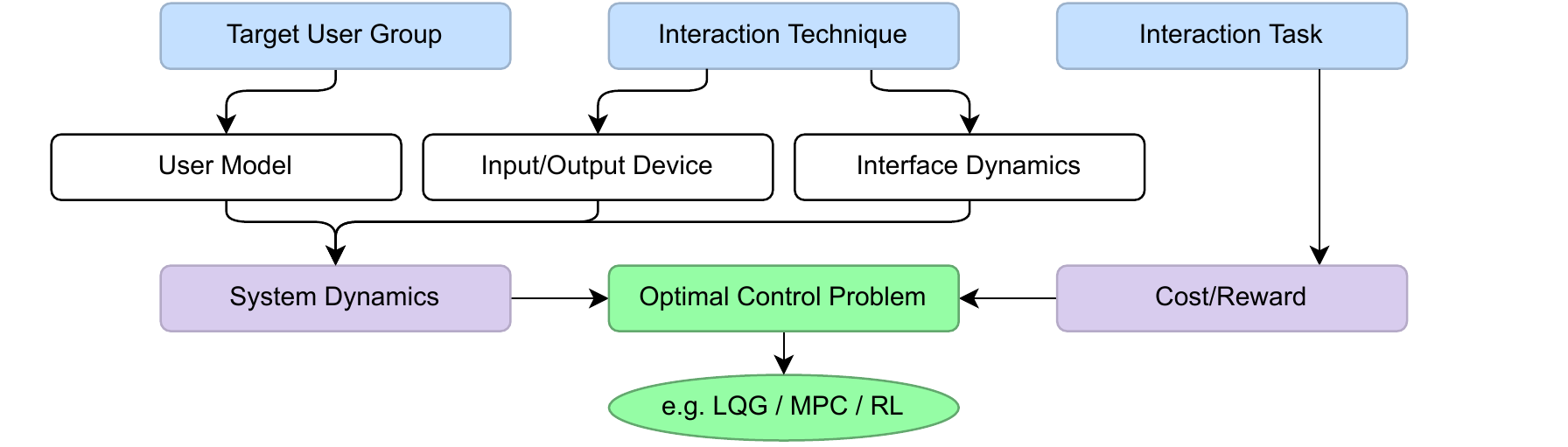}
	\caption{
	Our proposed optimization-based framework of human movement during interaction.
	The combination of system dynamics and cost/reward function results in an optimal control problem that can be solved using established methods.
	Figure adapted from~\cite{Klar2022}.
	}
	\label{fig:simulation_framework}
\end{figure}

In addition to developing novel displays and interaction techniques, simulation and optimization can be used to model human decision-making processes during interaction with computers and virtual systems.
A graphical representation of our proposed framework %
is given in Figure~\ref{fig:simulation_framework}.

Using the terminology of an \textit{Optimal Control Problem (OCP)}~\cite{Diedrichsen10}, we assume that for a given interaction task, humans aim to find a sequence of valid controls (e.g., neuromuscular control signals) such that the resulting movement minimizes a given internalized cost function reflecting both task-specific goals (e.g., pointing at a specific target) and their individual preferences (e.g., using the right index finger for pointing).
This is consistent with the general assumption of \textit{human rationality}, which states that given a set of options, humans will select the option that provides them with the greatest benefit~\cite{Silver21}.
One challenge is thus to ``encode'' a given task instruction, or rather, the internal objectives humans derive from that, into a mathematically precise cost function to be minimized (or, alternatively, a reward function to be maximized).

The human biomechanical model, input device, interface dynamics all have to be taken into account in this optimization.
These are included in the system dynamics, 
capturing all the constraints that stem from the user model (e.g., kinematics and perception), input and output devices (e.g., transfer functions), and interface dynamics.

Solving the resulting OCP leads to a simulation of the complete human-computer-loop that allows us to infer information such as movement times, cursor \textit{and} joint trajectories, or muscle expenditure.
Depending on the complexity of the system dynamics and cost function, the OCP can be solved using different methods. Three of the most important approaches to obtain (approximately) optimal movement trajectories for a given task, technique, and user group are discussed below.

\subsection{Linear-Quadratic Gaussian Regulator}
In the case of linear system dynamics and a quadratic cost function, the \textit{Linear-Quadratic Regulator (LQR)} yields a unique optimal control policy $\pi$, mapping an arbitrary state $x$ to the control $u^{\star}=\pi(x)$ that is optimal to apply when being at state $x$.
An extension to stochastic (linear) system dynamics, where controls and states are perturbed by Gaussian noise at each simulation step, is called the \textit{Linear-Quadratic Gaussian Regulator (LQG)}.

For both LQR and LQG, the control policy is again linear in the (expected) state, i.e., $\pi(x)=Lx$ holds for some matrix $L$ that can be computed once in advance (i.e., during the planning stage)~\cite{Todorov98}.
In particular, the generated solution trajectories are \textit{closed-loop}, i.e., the control adapts to perturbances that may occur during execution (e.g., due to model inaccuracies, system and control noise, or unexpected deviations from predicted states).
Since LQR and LQG allow to solve the OCP analytically, the optimal control policy can be computed very fast (typically within a few seconds) once before the movement starts, and optimal closed-loop controls are available in real-time during the movement.

We have introduced the LQR/LQG framework to the HCI community, and have shown its applicability to mouse pointing~\cite{Fischer20, Fischer22}.
For the LQG with control and observation noise, we have shown that effort costs applied on a finite time horizon, together with terminal distance and stability costs, are sufficient to generate characteristic cursor trajectories (e.g., bell-shaped velocity profiles, with speed-accuracy trade-off attributed to signal-dependent noise~\cite{HarrisWolpert98}). %
Moreover, reciprocal 1D mouse pointing trajectories were reproduced significantly better than using the minimum jerk model~\cite{Flash85} or pure dynamic models~\cite{Mueller17}, also capturing between-trial variability.

However, the assumption of linear dynamics prevents the LQG controller from dealing with nonlinearities, as they e.g., arise from the biomechanical constraints of human movements.
To overcome these limitations, we proposed the use of \textit{Model Predictive Control (MPC)}~\cite{Klar2022}.

\subsection{Model Predictive Control}\label{sec:MPC}

MPC is a receding horizon approach that handles \textit{nonlinearities}, provides optimality and convergence guarantees in many cases, and is inherently \textit{robust} to uncertainties due to its closed-loop nature~\cite{GP17}.
It has become a standard control method for (non)linear dynamical systems from both academic and application perspectives~\cite{QIN03}.
Instead of solving the original OCP that describes the entire interaction movement, a sequence of shorter OCPs is solved during motion execution.
Thus, the movement time does not have to be fixed in advance and, in addition, the complexity of the OCP that we need to solve is reduced in time.

External deviations that might occur during movement execution are taken into account by the receding horizon control principle. 
After solving an OCP, only the first part of the optimal control sequence is applied, the updated state of the system (which may deviate from the expected state) is observed, and a new OCP is setup starting from this state.
This procedure of alternating planning and execution steps is continued until the interaction task is completed.

The proposed MPC framework can be used with analytical (nonlinear) models or even ``black-box'' implementations.
As an application, we investigated mid-air pointing %
using a biomechanical model of the upper extremity implemented in the fast physics engine MuJoCo~\cite{mujoco}. 
This model is based on a state-of-the-art OpenSim model by Seth et al.~\cite{seth2018opensim}, consists of a torso, right shoulder, and arm, and has seven independent joints that can be directly actuated via applied torques. %
We combined this physical model with a second-order model for aggregated muscles at the individual joints derived from van der Helm et al.~\cite{van2000musculoskeletal}.
In order to assess how well our simulation reflects human movement, we captured motion data in a mid-air pointing user study featuring different pointing techniques.
By comparing three cost functions, we found that the combination of distance, control, and joint acceleration costs best explained the experimentally observed user behaviour in terms of both end-effector trajectories and joint angle sequences~\cite{Klar2022}.
We have also demonstrated
the ability to replicate the behavior of a \textit{specific} user, and to generate models of \textit{new} users by adjusting the model and/or control cost parameters.

\subsection{Model-Free Reinforcement Learning}
As an alternative approach to MPC, we have investigated the ability of \textit{model-free reinforcement learning (RL)} to simulate human movement during interaction.
While RL methods have a long tradition in robotics and character animation, in the last few years they have been increasingly used to predict human behavior and motion in areas such as neuroscience, digital health, and sports~~\cite{Bian20, Gottesman19, Liu22}.  %
In contrast to LQG, and similar to MPC, RL can handle complex, nonlinear system dynamics.
In contrast to MPC, the closed-loop policy learned by policy-gradient RL methods is not only (approximately) optimal \textit{for a single initial state}, but generalizes to \textit{arbitrary states} explored during training.
Practically, an RL policy thus needs to be ``trained'' only once (which, however, might take up to hours or even days) and can then be applied in real-time during execution (similar to LQG).
On the downside, there are much fewer theoretical optimality and stability guarantees for RL policies than for MPC and LQG, %
rendering RL approaches more ``experimental''.

Using the same state-of-the-art model of the upper extremity as in Section~\ref{sec:MPC} to simulate mid-air pointing movements, we have shown that an RL policy trained to minimize constant time costs only is able to generate movements that capture well-established characteristics. In particular, the simulation trajectories follow Fitts’ Law in the case of aimed reaching, and the \nicefrac{2}{3} Power Law for ellipse drawing, while generating bell-shaped velocity and N-shaped acceleration profiles~\cite{Fischer21}.

\begin{figure}
	\centering
	\includegraphics[width=0.24\linewidth, clip]{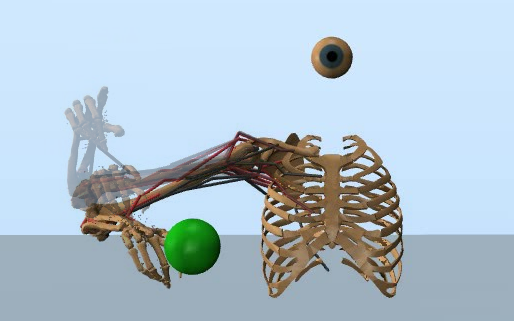}
	\hfill
	\includegraphics[width=0.24\linewidth, clip]{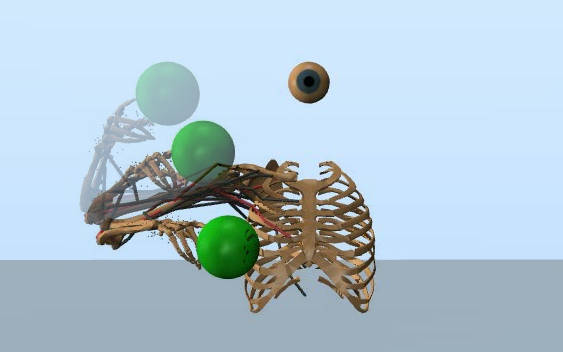}
	\hfill
	\includegraphics[width=0.24\linewidth, clip]{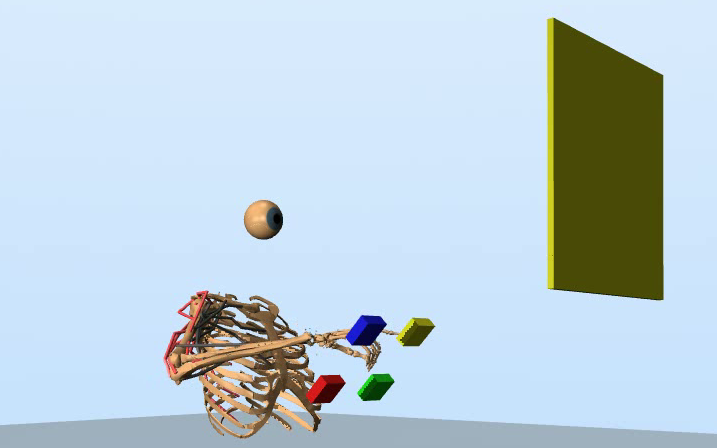}
	\hfill
	\includegraphics[width=0.24\linewidth, clip]{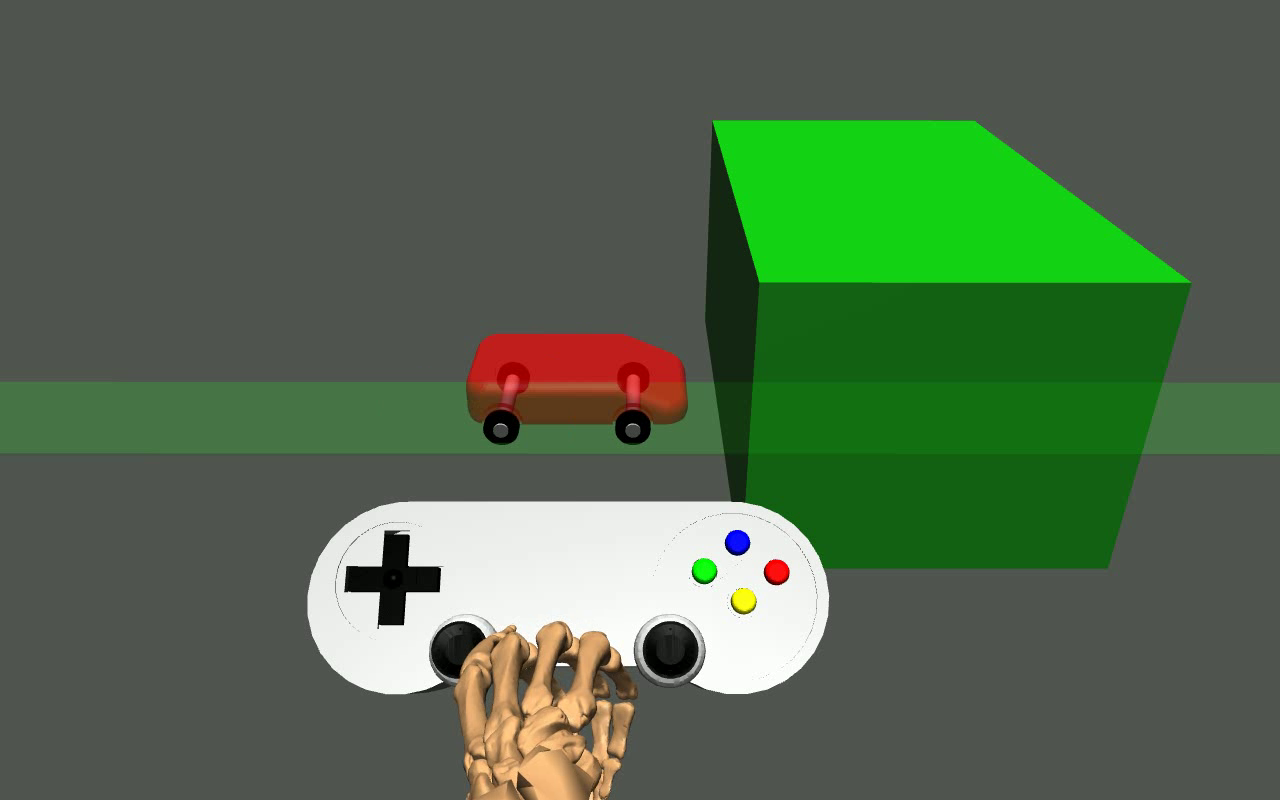}
	\caption{Using policies trained via RL, our simulation is capable of predicting motion in four interactive tasks with differing perceptual-motor requirements.
	Figure reprinted from~\cite{uitb}.
	}
	\label{fig:uitb-tasks}
\end{figure}

Building on these results, we have presented \textit{User-in-the-Box}~\cite{uitb}, a modular simulation framework that allows to combine a biomechanical model with one or multiple sensory input channels, an interaction task instance, and an RL method used to solve the resulting OCP. %
In this work we used a muscle-actuated MuJoCo version of the upper extremity model (with 5 shoulder and arm joints actuated via 26 muscles and tendons, and fixed torso and wrist), as well as RGB-D based vision, proprioceptive and haptic input channels, and task-specific reward functions. 
We simulated four movement-based interaction tasks of increasing difficulty: mid-air pointing, target tracking, choice reaction, and remote car control (see Figure~\ref{fig:uitb-tasks}).
Our trained models can successfully complete the respective interaction tasks, while exhibiting characteristic movement regularities such as Fitts' Law, showing that we are able to simulate interactive motion of real users.

\section{Discussion}\label{sec:discussion}
The approach presented here highlights the benefits that simulation and optimization can offer to the design, evaluation, and improvement of user interfaces.
However, in order to realize the full potential of model-based simulation, important aspects need to be discussed.

\emph{Model accuracy.} 
Although we are able to implement models of increasing complexity, the question is what aspects qualify for a ``good enough'' model of the human \textit{and} the interface, given a specific interaction task and technique.
Furthermore, what metrics are suitable and accepted by researchers and pracitioners alike to validate a model?

\emph{Generalizability.} 
Assuming we have created a ``good enough'' model for a specific interaction task and technique and went through the (painful) process of validating it, how easy is it to apply it to (slightly) different tasks or account for different user-specific characteristics? 
Do classes of cost/reward functions that encapsule a more general setting exist, or are we stuck with tuning these functions for each task, user model, and interaction technique?

\emph{Deployability.} 
We see a positive change within the HCI community of making code publicly available.
Building on this, what tools and tutorials specifically created for interface designers are necessary to wisely augment the user-centered design process with predictions from model-based simulations and allow for virtual prototyping? 
We believe that an ongoing, goal-oriented dialogue between researchers and practitioners is essential to ensure easy deployability and leverage the benefits of both real and simulated user data. 
Moreover, an interdisciplinary approach would help making virtual prototyping more accessible to different audiences.

\bibliographystyle{ACM-Reference-Format}
\bibliography{reference}

\end{document}